\documentclass[trackchanges,twocolumn]{aastex701}

\usepackage{graphicx}
\usepackage{microtype}
\usepackage{amsmath}
\usepackage{placeins}
\usepackage{txfonts}
\usepackage{xspace}
\usepackage{multirow}
\usepackage{xcolor}
\usepackage{url}
\usepackage{hyperref}
\usepackage[flushleft]{threeparttable}
\newcommand\nustar{\textit{NuSTAR}\xspace}
\newcommand\ixpe{\textit{IXPE}\xspace}

\begin{document} 

\title{The First X-ray Polarimetry of GRS 1739--278 Reveals Its Rapidly Spinning Black Hole}


\author[0000-0001-9893-8248]{Qing-Chang Zhao}
\affiliation{State Key Laboratory of Particle Astrophysics, Institute of High Energy Physics, Chinese Academy of Sciences, Beijing 100049, China} 
\affiliation{University of Chinese Academy of Sciences, Chinese Academy of Sciences, Beijing 100049, China}
\email{zhaoqc@ihep.ac.cn}

\author[0000-0003-0079-1239]{Michal Dovčiak}
\affiliation{Astronomical Institute of the Czech Academy of Sciences, Boční II 1401, 14100 Prague, Czech Republic}
\email{michal.dovciak@asu.cas.cz}
\author[0000-0002-5963-1494]{Hancheng Li}
\affiliation{Department of Astronomy, University of Geneva, 16 Chemin d’Ecogia, Versoix, CH-1290, Switzerland}
\email{Hancheng.Li@unige.ch}

\author[0000-0002-2705-4338]{Lian Tao}
\affiliation{State Key Laboratory of Particle Astrophysics, Institute of High Energy Physics, Chinese Academy of Sciences, Beijing 100049, China} 
\email{taolian@ihep.ac.cn}

\author[0000-0001-7584-6236]{Hua Feng}
\affiliation{State Key Laboratory of Particle Astrophysics, Institute of High Energy Physics, Chinese Academy of Sciences, Beijing 100049, China} 
\email{hfeng@ihep.ac.cn@ihep.ac.cn}

\author[0000-0002-1481-1870]{Federico Vincentelli}
\affiliation{Fluid and Complex Systems Centre, Coventry University, CV1 5FB, UK}
\affiliation{INAF—Istituto di Astrofisica e Planetologia Spaziali, Via del Fosso del Cavaliere 100, I-00133 Roma, Italy}
\affiliation{Department of Physics and Astronomy, University of Southampton, Highfield, Southampton, SO17 1BJ, UK}
\email{vincentelli.astro@gmail.com}

\author[0000-0002-2152-0916]{Giorgio Matt}
\affiliation{Dipartimento di Matematica e Fisica, Universit\`a degli Studi Roma Tre, via della Vasca Navale 84, 00146 Roma, Italy } 
\email{giorgio.matt@uniroma3.it}

\author[0000-0002-3638-0637]{Philip Kaaret}
\affiliation{NASA Marshall Space Flight Center, Huntsville, AL 35812, USA}
\email{philip.kaaret@nasa.gov}

\author[0000-0001-5586-1017]{Shuang-Nan Zhang}
\affiliation{State Key Laboratory of Particle Astrophysics, Institute of High Energy Physics, Chinese Academy of Sciences, Beijing 100049, China} 
\email{zhangsn@ihep.ac.cn}


\begin{abstract}
We present a joint spectro-polarimetric analysis of the black hole X-ray binary GRS 1739–278 during its 2025 mini-outburst, using simultaneous observations from \ixpe\ and \nustar. The \ixpe\ data show a polarization degree of ${\rm PD} = (2.3 \pm 0.4)$\% and polarization angle of ${\rm PA} = 62^{\circ} \pm 5^{\circ}$ in the 2$-$8 keV range. The model-independent analysis reveals that PD increases from $\sim2$\% at 2 keV to $\sim10$\% in the 6$-$8 keV band, while PA remains stable across the \ixpe\ band within statistical uncertainties. Broadband spectral modeling of the combined \ixpe\ and \nustar\ datasets shows that hard Comptonization contributes negligibly in this soft-state observation, while a substantial reflected component is required in addition to the thermal disk emission. We then model the \ixpe\ Stokes spectra using the {\tt kynbbrr} model. The best fits indicate that high-spin configurations enhance the contribution of the reflected returning radiation, which dominates the observed polarization properties. From the {\tt kynbbrr} modeling, we infer an extreme black hole spin of $a=0.994^{+0.004}_{-0.003}$ and a system inclination of $i = {54^\circ}^{+8^\circ}_{-4^\circ}$. Owing to the large contribution from the returning radiation, the observed polarization direction is nearly parallel to the projected system axis, the position angle of which is predicted at ${58^{\circ} \pm 4^{\circ}}$.
Our results demonstrate that X-ray polarimetry, combined with broadband spectroscopy, directly probes the geometry and relativistic effects in accretion disks around stellar-mass black holes.
\end{abstract}


%

\section{Introduction}

A black hole X-ray binary (BHXRB) consists of a stellar-mass black hole and a companion star. 
Most BHXRBs are transient, spending the majority of their time in quiescence. 
Occasionally, they undergo outbursts lasting from several weeks to months, during which their X-ray flux can increase by several orders of magnitude \citep{2006Remillard&McClintock,2016Belloni&Motta}. 
The X-ray emission of BHXRBs consists of multiple spectral components with both thermal and non-thermal origins. In the hard state, the spectrum exhibits a cutoff power-law shape, which is believed to originate from inverse Compton scattering of thermal seed photons from the accretion disk in the hot corona \citep{2005Markoff}. While in the soft state, the X-ray spectrum is predominated by a multi-temperature disk blackbody \citep{1973Shakura&Sunyaev}. In addition, 
BHXRBs often exhibit a reflection component in their X-ray spectra. 
This reflection component is generally produced when hard X-rays from the corona irradiate the accretion disk, giving rise to fluorescent lines and a Compton reflection hump—features commonly observed during the hard and intermediate states. 
Interestingly, several BHXRBs have also shown evidence of reflection in the soft state \citep{2020Connors,2021Connors,2021Lazar,2024Zhao}. 
In this regime, the corona is expected to be weak, and thus coronal illumination alone cannot fully account for the observed reflection features. 
A plausible explanation is returning radiation \citep{Cunningham1976}, in which a fraction of the thermal photons emitted by the disk are attracted by the black hole’s gravity and redirected back toward the disk surface, producing additional reflection signatures even in the absence of a strong corona. 

In the soft state, the accretion disk is generally believed to extend down to the innermost stable circular orbit (ISCO). 
The black hole spin determines the location of the ISCO and therefore influences the temperature of the inner disk, which shapes the observed thermal spectrum. 
Moreover, the black hole spin affects the observed X-ray polarization, as general relativistic rotation of the polarization plane near the black hole modifies the polarization degree (PD) and angle (PA) measured at infinity \citep{1980Connors}. Consequently, X-ray polarimetry provides an independent and powerful diagnostic of the spin and accretion geometry of black hole binaries. 
With the launch of the Imaging X-ray Polarimetry Explorer (\ixpe; \citealt{Weisskopf_ixpe}), this has now become observationally accessible thanks to \ixpe's high-sensitivity measurements of X-ray polarization in the 2--8~keV band.

A growing number of \ixpe\ observations have revealed diverse polarization behaviors among BHXRBs across different accretion states \citep[see][for a review]{2024Dovciak}.  
To date, about a dozen BHXRBs have been observed, including roughly ten during their soft states.  
Among these, five sources have shown significant detections of X-ray polarization (Cyg~X--1, \citealt{Steiner2024,Kravtsov2025}; Cyg~X--3, \citealt{Veledina2024b}; 4U 1630--47, \citealt{Ratheesh2024}; LMC X--3, \citealt{Svoboda2024}; and 4U 1957+115, \citealt{Marra2023}),
while the remaining observations have yielded only upper limits (LMC~X--1, \citealt{Podgorny2023}; Swift~J1727.8--1613, \citealt{Svoboda2024b}; Swift~J1518.0--5725, \citealt{Ling2024}; MAXI~J1744--294, \citealt{Marra2025}; and GX~339--4, \citealt{Mastroserio2024}).  
Notably, X-ray polarimetry of LMC~X--3 \citep{Svoboda2024}, 4U~1957+115 \citep{Marra2023} and Cyg~X--1 \citep{Steiner2024},  
has yielded direct spin constraints from polarization measurements: 
LMC~X--3 favors a slow spin with $a<0.7$, while both 4U~1957+115 and Cyg~X--1 imply a rapidly spinning black hole ($a\ge0.96$).  
These studies highlight the unique diagnostic power of X-ray polarimetry for probing black hole spin and accretion geometry, 
demonstrating its potential to complement traditional spectral and timing approaches.



GRS~1739$-$278 provides a valuable opportunity to extend such investigations.
This source was discovered in the Galactic bulge by \textit{SIGMA}/Granat during an outburst in 1996 \citep{Paul_GRS1739_Discovery,Vargas1997}. 
Its spectral and timing characteristics suggest that GRS~1739$-$278 is a black hole transient \citep{Borozdin1998,Borozdin2000_qpo,Wijnands2001_qpo,Mereminskiy2019}.
\citet{Greiner1996} estimated a distance of 6--8.5~kpc based on an extinction of $A_{\rm V}=14\pm2$~mag derived from the dust-scattering halo. 
The source underwent a major outburst in 2014 that lasted for more than one year \citep{Krimm_2014outburst,Filippova_2014outburst}. 
By modeling the reflection spectrum, \citet{Miller_2015} measured a high spin of $a_* = 0.8\pm0.2$ and an intermediate inclination of $43\fdg2\pm0\fdg5$. 
\citet{Draghis_2024} reported a similarly high spin of $a_* = 0.97^{+0.02}_{-0.07}$ but a higher inclination of $70^{\circ\,+5^\circ}_{\,-11^\circ}$. 
\citet{Wang2018} found that the inner disk radius reached the innermost stable circular orbit (ISCO) during the high state, placing an upper limit on the black hole mass of 18.3~$M_{\odot}$. 
By adopting the empirical transition luminosity of 1\%--4\% of $L_{\rm Edd}$, observed in black hole X-ray binaries during the change from the high-soft to the low-hard state, and assuming a distance of 6--8.5~kpc, and an inclination and spin as suggested by \citet{Miller_2015}, they further constrained the black hole mass to be in the range of 2.0--9.5~$M_{\odot}$. 
Following the major outburst in 2014, the source experienced at least two mini-outbursts \citep{Yan2017}. 
On 2025 September 8, \textit{Einstein Probe} (\textit{EP})/WXT reported that GRS~1739$-$278 was undergoing a new outburst, which was subsequently confirmed by follow-up observations with \textit{EP}/FXT \citep{Cheng2025}. 

In this paper, we report the first X-ray polarimetric measurements of GRS~1739$-$278 obtained by \ixpe during the soft state of its 2025 mini-outburst. The paper is organized as follows. In Section~\ref{sec:data_reduction}, we describe the observations and data reduction procedures.
The results are presented in Section~\ref{sec:results} and discussed in Section~\ref{sec:discussion}. Lastly, a conclusion and future prospects are given in Section~\ref{sec:conclusion}.


\begin{figure*}
    \centering
    \includegraphics[width=1.0\textwidth]{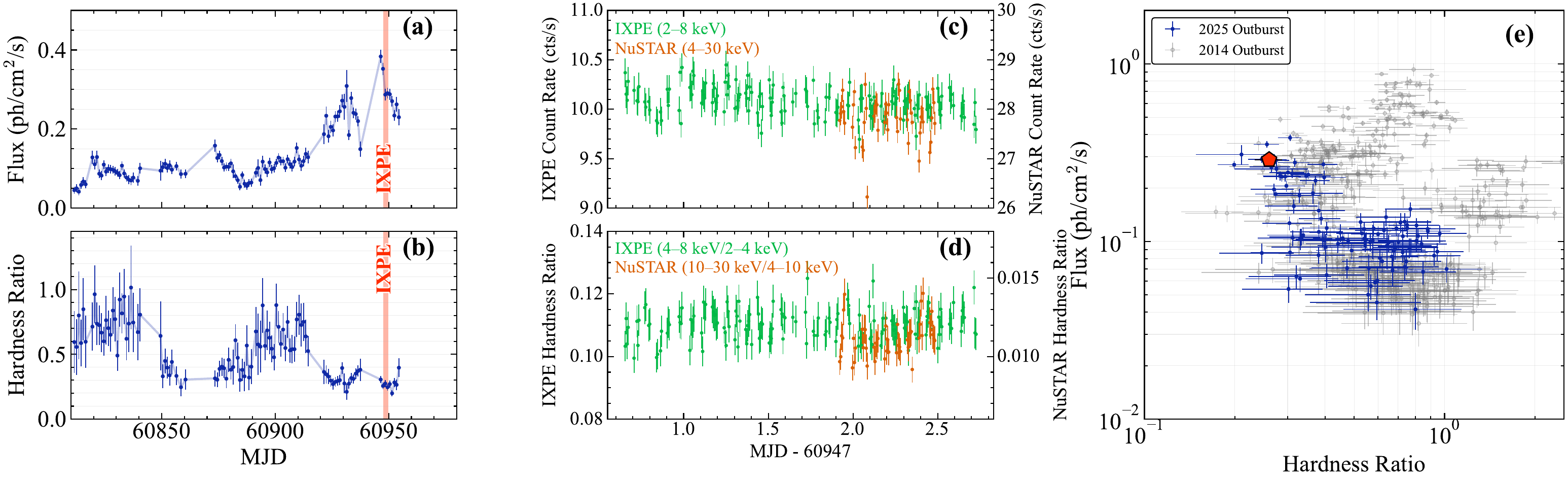}
    \caption{Evolution of GRS~1739$-$278 during the 2025 outburst. 
    (a) \textit{MAXI} light curve in the 2--20~keV band. 
    The time of the \ixpe\ observation is indicated by the shaded red strip. 
    (b) Same as panel~(a), but for the hardness ratio, defined as the count-rate ratio between 4--10~keV and 2--4~keV. 
    Panels (c) show the \ixpe\ (2--8~keV; green) and \nustar\ (4--30~keV; orange) count-rate light curves , respectively, computed with a bin size of 500 s. 
    Panels (d) present the corresponding hardness ratios, defined as the count-rate ratios between 4--8~keV and 2--4~keV for \ixpe, and between 10--30~keV and 4--10~keV for \nustar.
    (e) Hardness–intensity diagram (HID) of the 2025 outburst (blue) compared with that of the 2014 outburst (gray). 
    The red point marks the location of the \ixpe\ observation.}
    \label{fig:maxi_lc_hid}
\end{figure*}

\section{Observation and Data Reduction}
\label{sec:data_reduction}
\subsection{IXPE}
\ixpe \citep{Soffitta_etal_2021,Weisskopf_ixpe} observed GRS~1739$-$278 from 2025 September 29 to October 1 with a total exposure of 98.6~ks (ObsID~04251001, PI: Hancheng Li). 
This observation was carried out near the flux peak of the 2025 outburst, when the source exhibited a relatively soft hardness ratio, as shown in Fig.~\ref{fig:maxi_lc_hid} (a) and (b). As shown in panels (c) and (d), the count rate and hardness ratio of \ixpe are stable during the observation. 
In Fig.~\ref{fig:maxi_lc_hid} (e), we also show the hardness–intensity diagrams (HIDs) of the 2025 and 2014 outbursts for comparison. 
In 2014, the source underwent a major outburst, displaying a clear `q'-shaped loop in the HID. 
In contrast, during the 2025 outburst, the source did not trace such a `q'-shaped pattern.  
The \ixpe\ observation was conducted when the source was close to its softest state during this outburst. 

Our analysis starts from the level-2 data, which were further processed and analyzed using \texttt{ixpeobssim}~v31.0.3 \citep{Baldini_obssim} and \texttt{HEAsoft}~v6.35.2. 
During this observation, the data from DU2 were unavailable due to pixel failures that occurred in April 2025 (around MJD~60779). 
Consequently, the analysis is based solely on photons detected by DU1 and DU3.
We selected a circular region with a radius of 90$\arcsec$ for source extraction. Background subtraction was not performed as it was suggested to be unnecessary due to the high count rate of the source \citep{DiMarco_background}. We used \texttt{xpselect} to extract the photons within the source region in 2--8 keV. We performed an unweighted polarimetric analysis \citep{DiMarco2022}, and employed \texttt{ixpecalcarf} to generate the corresponding ARF/MRF files. The $I$, $Q$ and $U$ spectra were binned to 30 bins for spectro-polarimetric analysis, giving at least 380 counts per bin.

\subsection{NuSTAR}
\textit{NuSTAR} \citep{Harrison_NuSTAR} observed GRS~1739$-$278 on 2025 September 30 at 21:56:10 (ObsID~81160303004) with a total exposure time of approximately 21.7~ks (PI: Federico Vincentelli). 
Data reduction was performed using the \texttt{nupipeline} task from the \texttt{NuSTARDAS} package, distributed with \textsc{HEASoft}~v6.35.2, together with the calibration database (\texttt{CALDB}) version 20250729. 
Source events were extracted from circular regions with radii of 60\arcsec, while background events were extracted from regions of the same size located on the same detector but away from the source. 
Energy spectra and light curves were generated using the \texttt{nuproducts} task, and the spectra were grouped to have a minimum of 30 counts per bin. 
Because the background dominates above 30~keV, only the 4--30~keV energy range was used in our analysis. 

\section{Analysis and Results}
\label{sec:results}

\begin{figure*}
     \centering
    \includegraphics[width=0.75\textwidth]{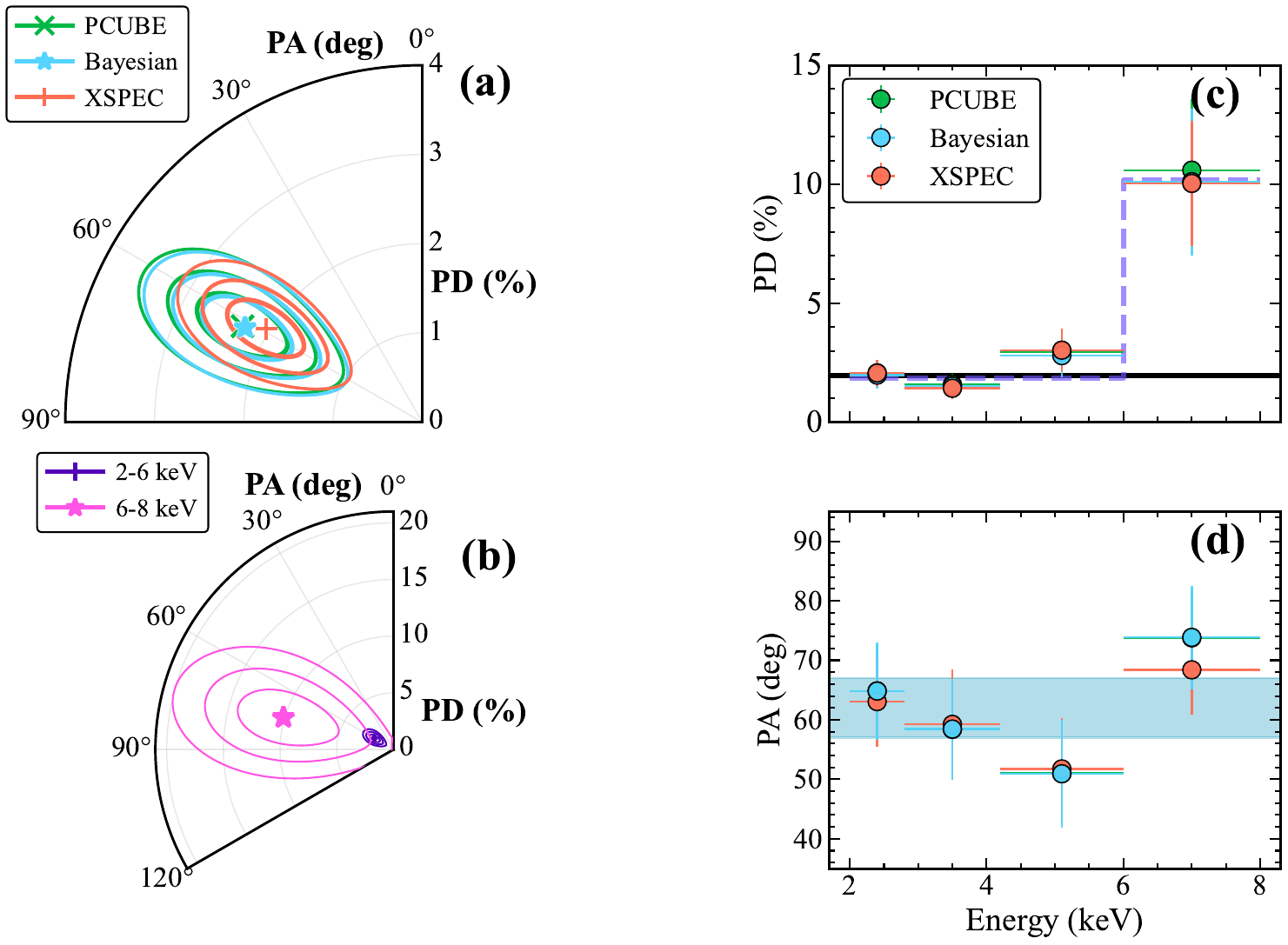}
    \caption{Polarization contours and energy dependence. (a) Polarization contours in the 2--8~keV band obtained with the \texttt{PCUBE} (green), \texttt{XSPEC} (orange), and Bayesian (cyan) methods, shown at the 1$\sigma$, 2$\sigma$, and 3$\sigma$ confidence levels. 
(b) Bayesian PA–PD contours for the 2–6~keV (purple) and 6–8~keV (pink) bands. 
(c) PD versus energy. The black solid and violet-blue dashed lines indicate the constant and bimodal fits, respectively. 
(d) PA versus energy, with the light-blue shaded region marking the energy-averaged PA (1$\sigma$ uncertainty).} 
    \label{fig:contour_polarization}
\end{figure*}

\subsection{Model-Independent Polarimetric Analysis}
We first carried out a model-independent weighted  (\texttt{NEEF}) polarimetric analysis using the \texttt{PCUBE} algorithm in \textsc{ixpeobssim}. 
This analysis yields a PD of $2.3\%\pm0.4\%$ ($>5\sigma$) and a PA of $62\degr\pm5\degr$. 
The corresponding statistical confidence contours are shown in Fig.~\ref{fig:contour_polarization}. 
We also applied a Bayesian approach, which provides a more robust estimate of the polarization parameters, particularly in the low-statistic regime \citep{maier2014,2025Li}. 
The posterior distribution is obtained using the following likelihood,
\begin{equation}
\begin{split}
P(p,\psi \mid p_{0},\psi_{0}) = 
\\
\frac{p\mu^{2}}{4\pi}\sqrt{\frac{N^{2}}{W_{2}}}
\exp\left[
-\frac{\mu^{2}N}{4}
\left(
p^{2} + p_{0}^{2}
- 2 p p_{0} \cos\left( 2(\psi - \psi_{0}) \right)
\right)
\right],
\end{split}
\end{equation}
with flat priors $p_{0}\in[0,1]$ and $\psi_{0}\in[0^\circ,180^\circ]$ with \texttt{NEFF} weighting. 
For the 2--8~keV band, this analysis yields a PD of $2.3\%\pm0.4\%$ and a PA of $62^\circ\pm5^\circ$, consistent with the \texttt{PCUBE} results.
We further investigated the energy dependence of the polarization using \texttt{PCUBE} and \texttt{Bayesian} methods to derive PD and PA in several energy intervals. 
The results are presented in panels (c) and (d) of Fig.~\ref{fig:contour_polarization}. 
The PD shows a possible increasing trend with energy. 
In particular, the highest-energy bin (6--8~keV) reaches $10\%\pm3\%$, significantly higher than the lower-energy values (see panels~(b) and (c) of Fig.~\ref{fig:contour_polarization}). 
To quantify this dependence, we fitted the PD--energy relation using a maximum-likelihood approach based on the probability distribution function from the Bayesian analysis. 
Two models were tested: (1) a constant PD model and (2) a bimodal model allowing different PDs below and above 6~keV. 
The best-fit results are shown in panel~(c) of Fig.~\ref{fig:contour_polarization}. 
The constant and bimodal fits are indicated by the black solid and violet-blue dashed lines, respectively. 
The Akaike and Bayesian Information Criteria (AIC = 16.66/12.31 and BIC = 16.04/11.08) favor the bimodal model. 
A $\chi^2$ test further supports this, giving $\chi^2/\mathrm{dof} = 10.56/3$ and a $p$-value of 0.0144, corresponding to a $\sim2.5\sigma$ deviation from the constant model. 
We also examined the PA as a function of energy using the same approach, but both constant and bimodal fits yielded comparable statistics (AIC = 30.69/29.44 and BIC = 29.44/28.21), indicating no significant energy dependence.


\subsection{Spectro-Polarimetric Analysis with simple polarization models}
\begin{figure}
    \centering
    \includegraphics[width=0.45\textwidth]{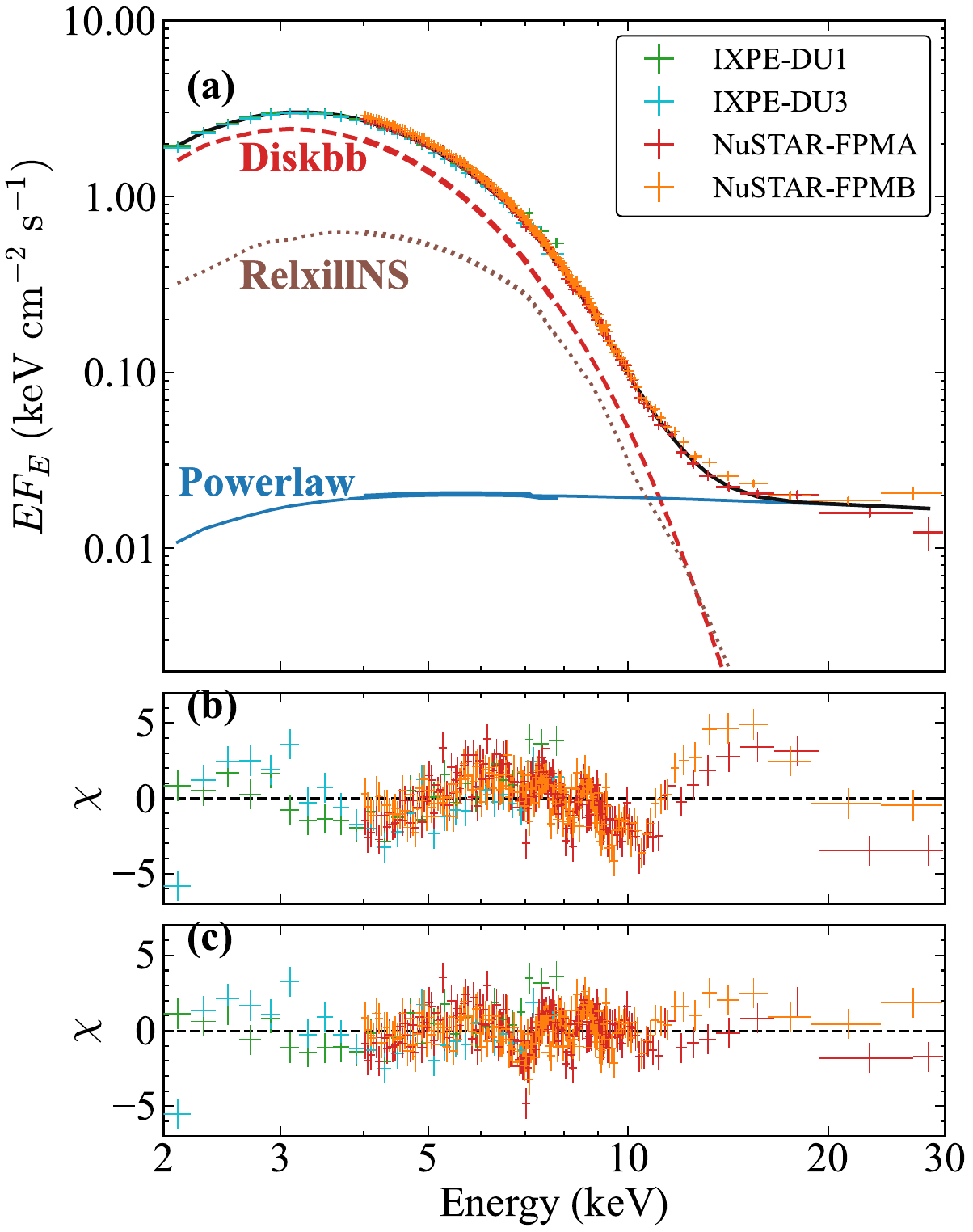}
    \caption{Spectral fit of the joint \ixpe\ and \nustar\ data with the model \texttt{constant*tbabs*(diskbb+powerlaw+relxillNS)}. 
(a) Unfolded spectra with individual model components shown. 
(b) Residuals from the fit without the \texttt{relxillNS} component. 
(c) Residuals for the best-fit model including \texttt{relxillNS}.
}
    \label{fig:ixpe_nustar_joint_fit}
\end{figure}

We also performed a spectro-polarimetric analysis using \ixpe and \nustar data with \textsc{XSPEC}~v12.15.0d. 
A \texttt{constant} component was included to account for cross-calibration differences between different instruments. 
Interstellar absorption was modeled using \texttt{tbabs} with elemental abundances from \citet{Wilms2000}. 
We first fitted the spectra with the model \texttt{constant*tbabs*(diskbb+powerlaw)}, 
which does not provide an adequate fit to the data, as shown in panel (b) of Fig.~\ref{fig:ixpe_nustar_joint_fit}.
The residuals reveal a reflection-like feature, likely produced by disk photons returning and being reflected off the disk surface \citep{Cunningham1976,Schnittman2009}. Since the source was observed in the soft state, where the contribution from hard Comptonization is weak, we included a \texttt{relxillNS} component in the model, whose incident spectrum is assumed to be a blackbody \citep{2022ApJ...926...13G}, rather than a cutoff power-law or Comptonized continuum as adopted in other \texttt{relxill} family models \citep{2014MNRAS.444L.100D,2014ApJ...782...76G}. 
In this model, we fixed the iron abundance at $A_{\rm Fe} = 1.5$, following \citet{2016ApJ...832..115F}. The spin is fixed at a extreme spin value of 0.998 \citep{Miller_2015,Draghis_2024}. The $kT_{\rm bb}$ was tied to the $T_{\rm in}$ of \texttt{diskbb}
. The inner disk radius was set to $R_{\rm ISCO}$ and the outer radius to $1000\,R_{\rm g}$. 
The parameters $\log N$, $\log \xi$, the inclination angle, and normalization were left free to vary. 
The best-fitting model, together with the corresponding residuals, is shown in panels (a) and (c) of Fig.~\ref{fig:ixpe_nustar_joint_fit}, respectively, and the best-fit parameters are listed in Table~\ref{tab:joint_spec_fit}. The inclination angle of the best fitting value is $45\degr\pm1\degr$, which is consistent with \citet{Miller_2015}.

\begin{table}
\centering
\caption{The best fitting parameters with constant*tbabs*(polconst*diskbb+polconst*powerlaw+polconst*relxillNS)}

\begin{tabular}{lccccc}
\hline\hline
Model Component & Parameter & Value \\
\hline

\texttt{tbabs} & $N_{\rm H}$ ($10^{22} \ \rm cm^{-2}$) & $3.25\pm0.03$ \\ \hline
\texttt{polconst} & PD (\%) & $1.8^{+2.0}_{-1.0}$ \\
& PA ($^\circ$) & $-35^{+44}_{-47}$ \\ 
\texttt{diskbb} & $T_{\rm in}$ (keV) & $0.932\pm0.006$ \\
& Normalization & $731\pm13$ \\ \hline
\texttt{polconst} & PD (\%) & $0^{\rm fixed}$ \\
& PA ($^\circ$) & $0^{\rm fixed}$ \\
\texttt{powerlaw} & $\Gamma$ & $2.16\pm0.08$ \\
& Normalization & $0.028^{+0.007}_{-0.005}$ \\ \hline
\texttt{polconst} & PD (\%) & $17^{+8}_{-7}$  \\ 
& PA ($^\circ$) & $57\pm12$ \\ 
\texttt{relxillNS} & Index1 & $3^{\rm fixed}$ \\
& Index2 & $3^{\rm fixed}$ \\
& $a$ & $0.998^{\rm fixed}$ \\
& $R_{\rm in}$ (ISCO) & $1^{\rm fixed}$ \\
& $R_{\rm out}$ ($R_{\rm g}$) & $1000^{\rm fixed}$ \\
& $kT_{\rm bb}$ & =$T_{\rm in}$ \\
& $\log \rm N$  & $\geq18.9$ \\
& $A_{\rm Fe}$  & $1.5^{\rm fixed}$ \\
& $\rm \log\xi$  & $2.38\pm0.05$ \\ 
& Inclination ($^\circ$)  & $45\pm1$ \\
& Normalization  & $0.0061\pm0.0006$ \\ \hline

Constant &IXPE (DU1) & $1.0^{\rm fixed}$ \\
& IXPE (DU3) & $0.985\pm0.001$ \\
& NuSTAR (FPMA) & $1.032\pm0.002$ \\
& NuSTAR (FPMB) & $1.069^{+0.002}_{-0.003}$ \\ \hline
& $\chi^2$/dof & $971.74 / 756$  \\
\hline\hline
\end{tabular}
\label{tab:joint_spec_fit}
\end{table}

After obtaining the best spectral model, we first estimated the overall polarization by applying a \texttt{polconst} component to the entire model. This fit yields a PD of $2.0\%\pm0.3\%$ and a PA of $59\degr\pm4\degr$. The corresponding confidence contours are shown in Fig.~\ref{fig:contour_polarization}(a). We then fixed the best-fitting spectral parameters and found the best-fitting polarimetric parameters in restricted energy intervals. The resulting energy-resolved polarization measurements are presented in Fig.~\ref{fig:contour_polarization}~(c)–(d). We then explored the polarization properties of individual spectral components by attaching separate \texttt{polconst} models to each of the \texttt{diskbb}, \texttt{powerlaw}, and \texttt{relxillNS} components. 
The PD and PA of the \texttt{powerlaw} component could not be constrained, likely due to its negligible flux contribution, and were therefore fixed to zero. 
The \texttt{diskbb} component shows a minor PD of $1.8^{+2.0}_{-1.0}$\,\%, 
while the \texttt{relxillNS} component exhibits a relatively high PD of $17^{+8}_{-7}$\,\%. The PAs of the two components are approximately orthogonal, consistent with the theoretical expectation.

\subsection{Spectro-Polarimetric Analysis with Novikov-Thorne standard disk model and reflection of disk self-irradiation}

The previous spectro-polarimetric fits using \texttt{relxillNS} with simple polarization components (\texttt{polconst}) are empirical, since \texttt{relxillNS} currently neither predicts intrinsic polarization nor links the reflected emission self-consistently to that of the thermal disk. 
To physically interpret the measured polarization and its energy dependence, 
we therefore employ the \texttt{kynbbrr} model \citep{Mikusincova2023, Taverna2020, Dovciak2008}, which treats the direct disk emission and the returning radiation together and computes their polarization in a general relativistic framework.
It is important to note that \texttt{relxillNS} and \texttt{kynbbrr} are complementary: \texttt{relxillNS} offers a more detailed description of the reflection spectrum but lacks intrinsic polarimetry, whereas \texttt{kynbbrr} includes polarization self-consistently but employs a simplified reflection prescription. These capabilities are expected to converge once the blackbody reflection tables of \citet{Podgorny2025} and the improved thermal-disk polarization treatment of \citet{Marra2025b} are incorporated into \texttt{kynbbrr} in the future.

\begin{figure}
\hspace*{5mm}{\bf (a)}\\[-4mm]
\hspace*{1cm}\includegraphics[width=0.3\textwidth]{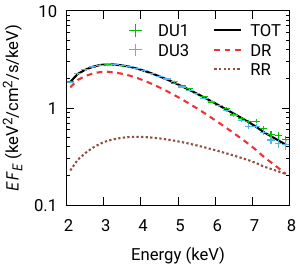}\\[1mm]
\hspace*{5mm}{\bf (b)}\\[-4mm]
\hspace*{1cm}\includegraphics[width=0.3\textwidth]{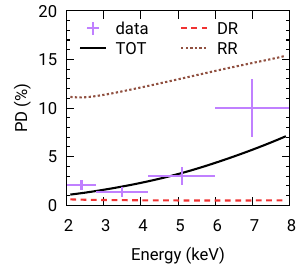}\\[1mm]
\hspace*{5mm}{\bf (c)}\\[-4mm]
\hspace*{1cm}\includegraphics[width=0.3\textwidth]{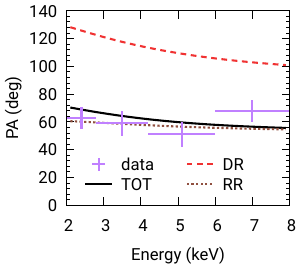}
\caption{{\bf a)} The decomposition of the best-fit-model total flux (TOT) into the direct thermal radiation (DR) and reflected radiation (RR) due to the disk self-irradiation given by the {\tt kynbbrr} model. 
{\bf b)} Similar to a) but for the PD.
{\bf c)} Similar to b) but for the PA.
}
\label{fig:kynbbrr_fit}
\end{figure}

The {\tt kynbbrr} model assumes a Novikov–Thorne accretion disk \citep{Novikov1973, Page1974}, with polarization approximated following \citet{Chandrasekhar1960} for Thomson scattering in a semi-infinite electron atmosphere.
The energy shift due to Comptonization is treated through a constant color-correction factor, $f_{\rm col}$ \citep[e.g.][]{Shimura1995, Davis2005}.
Here, we adopt a typical value for black hole X-ray binaries of $f_{\rm col}\!=\!1.7$ \citep{Davis2005}. The reflection of returning radiation is included and assumed to be polarized according to the multiple-scattering approximation of \citet{Chandrasekhar1960} for reflection from a cold electron atmosphere. In {\tt kynbbrr} model, the strength of the reflection component is parameterized by the energy-independent albedo, $A$, where $A\!=\!0$ corresponds to no reflection and $A\!=\!1$ to full reflection.

\begin{table}
\centering
\caption{Parameters of the {\tt const*tbabs(kynbbrr)} model. The errors are given at 1-$\sigma$ significance. The errors on albedo, $A$, spin, $a$, and inclination, $i$, were estimated from the contour graphs, see Fig.~\ref{fig:spin_inclination_contour}, the rest of the errors are 1-$\sigma$ statistical uncertainties from the fit.}
\begin{tabular}{lccccc}
\hline\hline
Model Component & Parameter & Value \\
\hline
\texttt{tbabs}   & $N_{\rm H}$ ($10^{22} \ \rm cm^{-2}$) & $2.20\pm0.04$ \\ \hline
\texttt{kynbbrr} & $A$ & $1^{+0}_{-0.3}$ \\
& $f_{\rm col}$ & $1.7^{\rm fixed}$  \\
& $a$ & $0.994^{+0.004}_{-0.003}$\\
& $M_{\bullet}$ ($M_\odot$) & $16\pm1$ \\
& $\dot{M}$ ($\dot{M}_{\rm Edd}$) & $0.043\pm0.004$  \\
& $i$ ($^\circ$) & $54^{+8}_{-4}$ \\
& $\psi$ ($^\circ$) & $58 \pm 4$ \\
& Normalization  & $2.03$ ($D=7\,$kpc)\\ \hline
constant &IXPE (DU1) & $1.0^{\rm fixed}$ \\
& IXPE (DU3) & $0.999\pm0.001$ \\
\hline
& $\chi^2$/dof & $213.82 / 171$  \\
\hline\hline
\end{tabular}
\label{tab:kynbbrr_fit}
\end{table}

In our analysis, we fitted for the albedo ($A$), BH spin ($a$), BH mass ($M_\bullet$), accretion rate ($\dot{M}$), system inclination ($i$, angle between the line of sight and the system axis), and the orientation of the system axis on the sky ($\psi$, measured from north through east). 
Because the BH mass, accretion rate, and distance ($D$), are mutually degenerate when fitting the thermal disk emission, we first fixed the distance\footnote{In {\tt kynbbrr}, the distance is encoded in the model normalization as $1/D^2$, where $D$ is expressed in units of 10\,kpc.} at 7\,kpc. Subsequently, we also allowed the distance to vary to test whether an improved fit could be obtained; as expected, its best-fit value changed only slightly.
In spectral fitting of thermal disk emission, the BH spin and the system inclination are also partly degenerate with the parameters mentioned above. However, since the polarization properties depend strongly on both spin and inclination, these quantities can be constrained more robustly when fitting all three Stokes parameters provided by the \ixpe observations.
In contrast, the orientation of the system axis does not affect the spectral shape and is constrained solely by the observed polarization properties. 
The best-fit values of all model parameters are summarized in Table~\ref{tab:kynbbrr_fit}. The fit was performed using only the \ixpe data, binned into 30 bins, and yields $\chi^2/$dof$=\!213.82/171$. 
A more detailed description of the analysis using the {\tt kynbbrr} model is provided in Appendix~\ref{app:kynbbrr} of the Supplementary Material.

The best-fit model spectrum, PD, and PA are shown in Fig.~\ref{fig:kynbbrr_fit}, with the contributions from each of the two components, the directly observed thermal emission and reflected returning radiation, depicted separately. While the direct flux (contributing $\sim\!80\%$) is about four times higher than the reflected one ($\sim\!20\%$), its PD ($\sim\!0.5\%$, almost independent of energy) is much lower than that of the reflected radiation. Furthermore, both the flux contribution and the PD of the reflected component increase with energy --- the former from 14\% in the 2–3\,keV band to 45.6\% in 7–8\,keV, and the latter almost linearly from 11\% at 2\,keV to 15\% at 8\,keV.
The reflected component therefore dominates the polarization properties in the \ixpe\ energy band, and its relative importance increases with energy.
Note that the direction of polarization of the direct thermal radiation is misaligned by $\sim\!40^\circ\text{ at 8\,keV and }\sim\!70^\circ$ at 2\,keV with respect to the polarization direction of the reflected radiation.

\section{Discussion}
\label{sec:discussion}

\subsection{X-ray polarization of GRS~1739--278 measured by \ixpe}
We report measurement of a significant PD of $2.3\%\pm0.4\%$ and a PA of $62\degr\pm5\degr$ in the 2--8~keV band measurements during the soft state of GRS~1739--278 in its 2025 mini-outburst. However, since the position angle of the radio jet is not available at the time of writing, we are unable to compare the X-ray PA with the jet orientation.

The model-independent analysis, as shown in Fig.~\ref{fig:contour_polarization}, reveals an increasing trend of PD with photon energy. As the bimodal model is preferred over the constant model, the PD at the highest-energy bin, 6-8 keV ($\sim$10\%), exceeds that at 2-6 keV ($\sim$2\%). 
The $\sim$10\% PD in the 6--8~keV band is unlikely to originate from direct disk emission alone. 
A standard thin disk produces only a few percent PD (even at high inclination), and additionally, relativistic effects would further suppress it. 
Thus, for a system viewed at moderate inclination, the observed high PD indicates the presence of an additional polarized component.



\subsection{Returning radiation revealed by \nustar}

We performed an empirical joint spectro-polarimetric analysis using \ixpe\ and \nustar. The broad \nustar\ coverage reveals clear reflection features. Accordingly, adding \texttt{relxillNS} to the \texttt{diskbb} + \texttt{powerlaw} continuum significantly improves the fit, while the \texttt{powerlaw} contributes only a negligible fraction of the total flux. The empirical spectro-polarimetric decomposition further indicates that the \texttt{diskbb} component carries only a low PD of $1.8^{+2.0}_{-1.0}$\,\%, whereas the \texttt{relxillNS} component exhibits a substantially higher PD of $17^{+8}_{-7}$\,\%, with their PAs being approximately orthogonal.

Since a standard thin disk at moderate inclination can produce only a few percent polarization, the $\sim$10\% PD observed in the 6–8~keV band cannot originate from the direct thermal emission alone. Instead, this behavior is naturally explained if returning radiation contributes increasingly at higher energies, where its intrinsically higher polarization dominates the observed signal. While the \texttt{polconst} decomposition supports this interpretation, degeneracies limit the physical robustness of the empirical model. We therefore adopted the more self-consistent \texttt{kynbbrr} model.

\subsection{Spin and inclination constraints from {\tt kynbbrr}}
\label{sec:4.3}
\begin{figure}
{\bf (a)}\hspace*{3.9cm}{\bf (b)}\\[-2mm]
\hspace*{-1mm}
\includegraphics[width=0.236\textwidth]{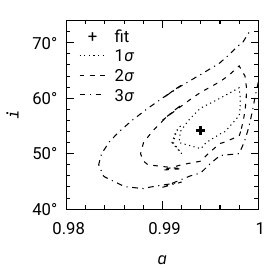}
\hspace*{-1mm}
\includegraphics[width=0.236\textwidth]{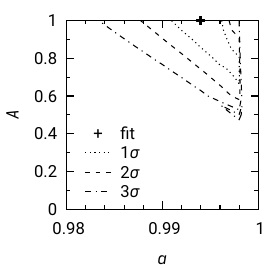}
\caption{(a) The spin--inclination and (b) the spin--albedo contour graphs produced with the {\tt XSPEC} {\tt steppar} command. The 1-, 2- and 3-$\sigma$ values are denoted by dotted, dashed and dash-dotted lines, respectively, while the best fit is shown by a plus sign.
}
\label{fig:spin_inclination_contour}
\end{figure}

The {\tt kynbbrr} results qualitatively confirm those obtained using simpler polarization models, while providing new constraints on the physical parameters of the system. As shown in Fig.~\ref{fig:spin_inclination_contour}, spin and inclination are stringently constrained at $a=0.994^{+0.004}_{-0.003}$ and $i={54^\circ}^{+8^\circ}_{-4^\circ}$, respectively. A large spin is also consistent with the required high contribution of returning radiation (high reflectivity, with an albedo of $A=1^{+0}_{-0.3}$, see 
the same figure).
In addition, for the first time, {\tt kynbbrr} also predicts the orientation of the system axis on the sky, $\psi=58^\circ\pm4^\circ$, which could be confirmed by future radio observations.

Our spin measurement from polarimetry is fully consistent (within the 1$\sigma$ statistical uncertainties) with the estimates obtained from reflection modeling in the hard state of this source by \cite{Miller_2015} ($a\!=\!0.8 \pm 0.2$) and \cite{Draghis_2024} ($a\!=\!0.97^{+0.02}_{-0.07}$), while our inclination lies between those reported in these works ($43\fdg2 \pm 0\fdg5$ and $70^{\circ\,+5^\circ}_{\,-11^\circ}$, respectively). Assuming a distance of 7\,kpc (i.e. within the range 6–8.5\,kpc; \citealt{Greiner1996}), we obtain an accretion rate of $\sim\!4$\%\,$\dot{M}_{\rm Edd}$ and a black hole mass of $\sim\!16\,M_{\odot}$, which is consistent with the upper limit of $18.3\,M_{\odot}$ estimated by \cite{Wang2018} from the 2014 soft state. Our result differs from their more stringent upper limit of $9.5\,M_{\odot}$, likely because of their slightly lower inclination adopted from \cite{Miller_2015}, and because their modeling did not account for reflection due to disk self-irradiation.

\subsection{Other physical scenarios}

The large reflection fraction of the returning radiation could be alternatively caused by the accretion disk extending appreciably vertically from the equatorial plane, as expected in slim or geometrically thick disks at high accretion rates \citep{Sadowski2011}. Such configurations are thought to arise in the soft state when $\dot{M}\!\sim\!\dot{M}_{\rm Edd}$, whereas in our case the inferred accretion rate is low ($\sim\!0.04\,\dot{M}_{\rm Edd}$), making a slim/thick inner disk less likely. We also note that the PD was found to remain below $\sim\!$10\% for disk aspect ratios $H/R\!\lesssim\!0.3$ \citep{West2023}.

The net polarization observed in GRS~1739$-$278 could, in principle, arise in a low-temperature ($kT_{\rm e}\!\lesssim\!10$\,keV) electron-scattering atmosphere above the accretion disk. 
In such a layer, photons experience zero, single, or multiple scatterings depending on its optical depth \citep[e.g.][]{Poutanen1993,Poutanen1996}. Because unscattered and singly scattered photons are polarized parallel to the disk plane, whereas multiply scattered photons are polarized perpendicularly, the net polarization remains low and cannot reproduce the high polarization observed in the 6–8\,keV band.

Disk winds could contribute appreciably to the X-ray polarization signal \citep[e.g.][]{Veledina2024,Nitindala2025}.
However, neither current nor archival observations of GRS~1739$-$278 show convincing signatures of scattering/reflective disk-wind material, which would be expected unless the wind were fully ionized throughout (an unlikely situation). Moreover, a purely wind-induced origin has not been shown to reproduce the pronounced rise of the PD with energy observed here.

\subsection{Comparison with other \ixpe soft-state observations of BH XRBs}

Among soft-state BH XRBs observed by \ixpe, GRS~1739$-$278 provides one of the clearest cases in which the polarized emission originates almost purely from the standard accretion disk and its self-irradiation.
In Cyg X$-$1, where polarimetric observations also confirm an extremely high black-hole spin \citep{Steiner2024}, the soft-state spectrum contains a significant Comptonized component and its reflection from the disk, both of which dilute the intrinsic polarization of the reflected returning thermal radiation, especially above 5\,keV.
Although less prominent, the Comptonized component contributing to the \ixpe polarization signal observed in 4U~1957$+$11 \citep{Marra2023} was still stronger than in GRS~1739$-$278. The \ixpe observation of 4U~1957$+$11 likewise favored very high spin and albedo values (see Figs.~7, 9, and 10 in \citealt{Marra2023}) and required an inclination of $\gtrsim50^\circ$.
The \ixpe polarimetric data of LMC~X$-$3 \citep{Svoboda2024} are consistent with a low black-hole spin and low albedo, although the energy dependence of its PD appears qualitatively similar to that observed in GRS~1739$-$278, albeit with larger uncertainties (see our Fig.~\ref{fig:kynbbrr_fit} and Fig.~2 in \citealt{Svoboda2024}). Since this source has more prior constraints than GRS~1739$-$278 --- its mass, inclination, and distance are all well determined --- the apparent low-spin fit may be driven by limited statistics that still allow an energy-independent PD. If a rise of the PD with energy is confirmed in future observations, the standard disk model would no longer reproduce the data, and similar polarization behaviour could then arise from mechanisms other than disk self-irradiation. Nevertheless, as LMC~X$-$3 accretes at an order-of-magnitude higher rate ($\sim\!40$\% $\dot{M}_{\rm Edd}$ compared to $\sim\!4$\% $\dot{M}_{\rm Edd}$), the same physical interpretation may not be directly applicable to GRS~1739$-$278.\newline

\section{Conclusions and future prospects}
\label{sec:conclusion}

In this paper, we report the first detection of X-ray polarization from GRS~1739$-$278. 
We measure a PD of $2.3\%\pm0.4\%$ ($>5\sigma$) at a PA of $62^{\circ}\pm5^{\circ}$ in the soft state. 
The PD increases with energy, reaching $\sim$10\% in the 6--8\,keV band, which cannot be explained by direct disk emission alone at the system’s moderate inclination. 
A joint spectral–polarimetric analysis indicates that the polarized signal is dominated by returning radiation, where thermal disk photons are gravitationally bent back to and reprocessed by the disk surface. 
By modeling the direct and returning disk emission self-consistently with \texttt{kynbbrr}, we obtain tight constraints on the black hole parameters, 
inferring a rapidly spinning black hole ($a = 0.994^{+0.004}_{-0.003}$) viewed at an inclination of $i = 54^{\circ}{}^{+8^\circ}_{-4^\circ}$. 
This work demonstrates that X-ray polarimetry provides a direct and powerful probe of the innermost accretion geometry around stellar-mass black holes.

Future X-ray polarimetric observations of GRS~1739$-$278 and similar systems will be crucial to further test the role of disk self-irradiation in rapidly spinning black holes.
The forthcoming {\it eXTP} mission will combine larger effective area with fast timing capabilities, enabling time-resolved polarization studies that can track changes in the disk and coronal geometry and yield tighter spin–inclination constraints in BH XRBs \citep{Zhang2019,Bu2025}.
The proposed ESA {\it EXPO} mission (an evolution of the {\it NGXP} concept by \citealt{Soffitta2021}), currently under consideration within the M8 call, would extend \ixpe’s capabilities to higher photon energies and larger effective area, enabling much more detailed X-ray polarimetry of both transient and persistent sources and providing improved diagnostics of black-hole accretion-flow geometry and system properties.

\begin{acknowledgements}

We thank the anonymous referee for the constructive comments that helped improve the manuscript. We acknowledge funding support from the National Natural Science Foundation of China under grants Nos.\ 12122306, 12025301, \& 12103027, and the Strategic Priority Research Program of the Chinese Academy of Sciences. MD thanks GACR project 21-06825X for the support and institutional support from RVO:67985815. GM acknowledges financial support from grants: PRIN 2022 - 2022LWPEXW - “An X-ray view of compact objects in polarized light”, CUP C53D23001180006; ASI-INAF-2022-19-HH.0.

We thank the \ixpe team for promptly scheduling the DDT observation and the \nustar team for coordinating the joint observation. 
This work reports observations obtained with the Imaging X-ray Polarimetry Explorer (IXPE), a joint US (NASA) and Italian (ASI) mission, led by Marshall Space Flight Center (MSFC). The research uses data products provided by the IXPE Science Operations Center (MSFC), using algorithms developed by the IXPE Collaboration (MSFC, Istituto Nazionale di Astrofisica - INAF, Istituto Nazionale di Fisica Nucleare - INFN, ASI Space Science Data Center - SSDC), and distributed by the High-Energy Astrophysics Science Archive Research Center (HEASARC).
\end{acknowledgements}

\appendix

\counterwithin{table}{section}
\counterwithin{figure}{section}

\section{Supplementary material}

\subsection{Details on fitting with {\tt kynbbrr} model.}
\label{app:kynbbrr}

The basic description of the fitting procedure with \texttt{kynbbrr} is provided in the main text. Here, we present additional details on the modelling approach and fitting methodology.

\begin{figure}
\hspace*{7mm}{\bf (a)}\hspace*{7.9cm}{\bf (b)}\\[-1mm]
\hspace*{1cm}\includegraphics[width=0.4\textwidth]{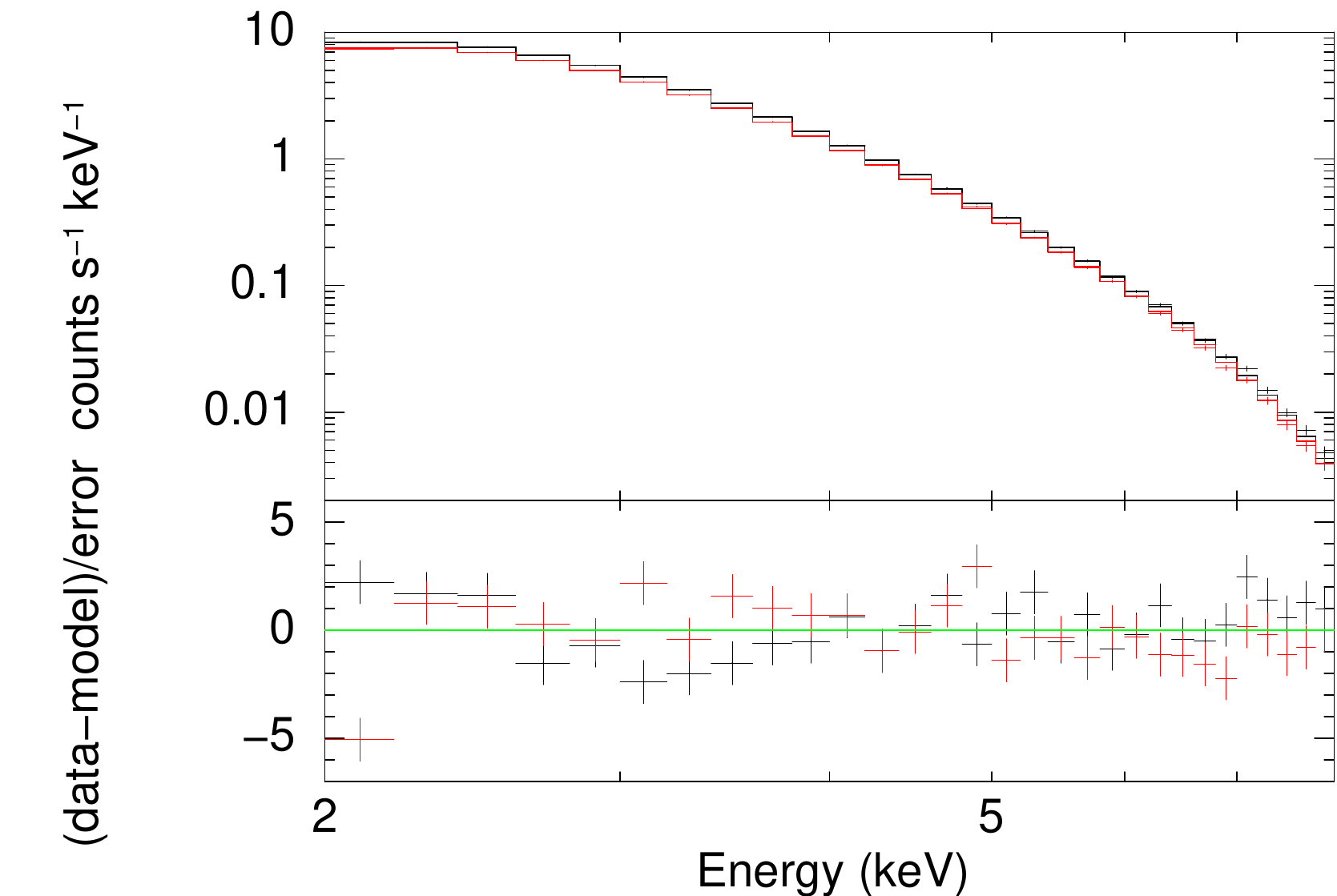}
\hspace*{1cm}\includegraphics[width=0.4\textwidth]{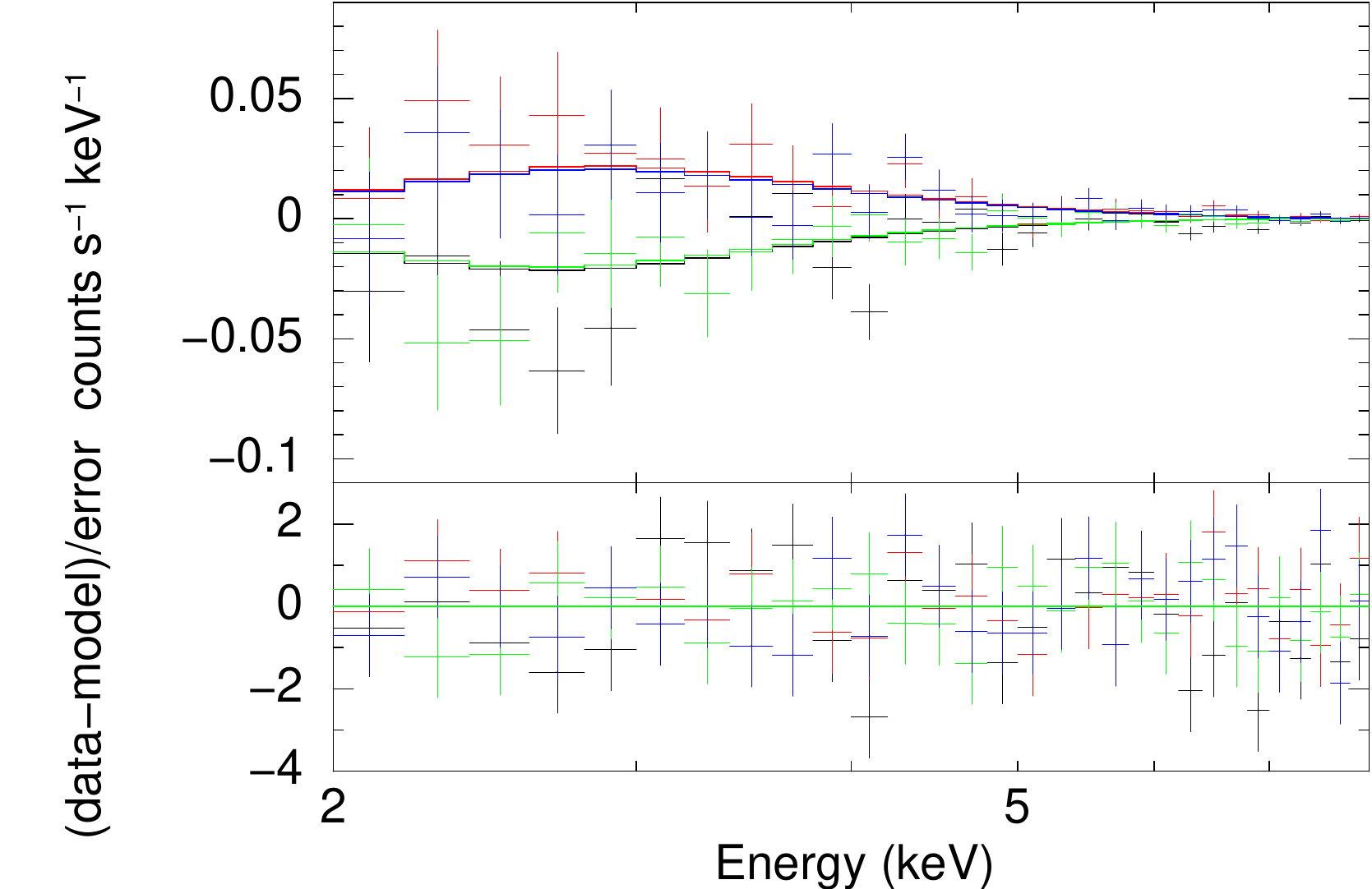}
\caption{{\bf (a)} The countrate, $i$, for the best fit of the data binned in 30 bins with {\tt kynbbrr} model (top panel), and the $\Delta\chi^2-$residuals between the data and model normalized by the statistical uncertainty (bottom panel). The data and model is denoted by black for DU1 and red for DU3. {\bf (b)} Similar to a) but for the specific Stokes parameters $q$ denoted by black (DU1) and green (DU3), and  $u$ denoted by blue (DU3) and red (DU1).}
\label{fig:kynbbrr_fit_stokes}
\end{figure}

\begin{figure}
{\bf (a)}\hspace*{4.2cm}{\bf (b)}\hspace*{4.2cm}{\bf (c)}
\\[-1mm]
\hspace*{-1mm}
\includegraphics[width=0.23\textwidth]{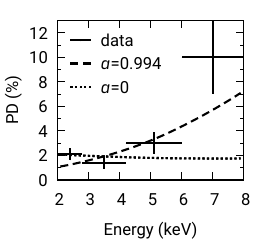}
\hspace*{3mm}
\includegraphics[width=0.23\textwidth]{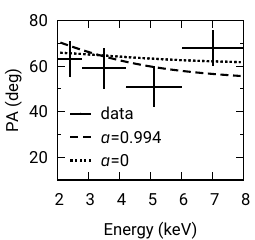}
\hspace*{5mm}
\includegraphics[width=0.476\textwidth]{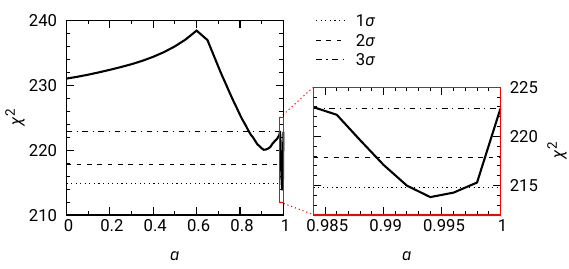}
\caption{{\bf (a)} The comparison of the observed PD (points) with the best fit model with the BH spin $a=0.994$ and disk albedo $A=1$ (dashed) and the model with the BH spin $a=0$ and disk albedo $A=0$ (dotted). Note that the latter model is not able to fit the high energy PD behaviour. 
{\bf (b)} Same as (a) but for the observed PA.
{\bf (c)} The $\chi^2$ dependence on the BH spin values using \texttt{const*tbabs(kynbbrr)} model. The best fit is reached for BH spin of $a=0.994$. The zoom-in region for the highest spin values is shown within the red frame.
}
\label{fig:kynbbrr_fit2}
\end{figure}

We employed the full model \texttt{const*tbabs(kynbbrr)}. As stated in the main text, the BH spin was among the fitted parameters. However, the current version of the \texttt{kynbbrr} model does not support arbitrary spin values, the spin parameter is restricted to a predefined grid: 
$a=0,\,$$0.05,\,$\dots,\,$0.65,\,$$0.7,\,$$0.71,\,$\dots,\,$0.97,\,$
$0.98,\,$$0.982,\,$\dots,\,$0.998,\,$$1$. The best-fit spin value was determined using the {\tt XSPEC steppar} command to evaluate the $\chi^2$ dependence over this grid (see panel (c) in Fig~\ref{fig:kynbbrr_fit2}). The corresponding uncertainty was obtained from contour plots of $\chi^2$ generated with \texttt{steppar} for the spin-inclination and spin-albedo parameter pairs (see Fig.~\ref{fig:spin_inclination_contour}). The best fit is shown in Fig.~\ref{fig:kynbbrr_fit_stokes} for all Stokes parameters, including their residuals between the data and the model. For the \ixpe data binned into 30 bins, the fit yields $\chi^2/$dof$=\!213.82/171$, corresponding to a null-hypothesis probability of $1.66\%$. 
Using the same best-fit parameters to evaluate the fit quality for the original unbinned \ixpe\ data gives $\chi^2/$dof$=939.75/891$, corresponding to a null-hypothesis probability of 13\%. 

The spin dependence of $\chi^2$ is shown in panel (c) of Fig.~\ref{fig:kynbbrr_fit2}. The best fit is achieved for an extreme BH spin, with the corresponding parameter values listed in Table~\ref{tab:kynbbrr_fit}. A high reflection fraction of returning radiation (albedo) is required, and the system axis roughly coincides with the observed polarization direction ($\psi\!\sim\!58^\circ$). The fit for a non-rotating black hole ($a\!=\!0$) is worse by $\Delta\chi^2\!=\!17.2$, corresponding to a significance of $\sim\!4\,\sigma$ for one parameter. The low-spin fit required an albedo of $A\!=\!0$, a higher inclination of $\sim\!65^\circ$, and a system-axis orientation almost perpendicular to the observed polarization direction ($\psi\!\sim\!-23^\circ$). Intermediate to high spins ($0.5\lesssim\!a\lesssim\!0.98$) required even higher inclinations, approaching an edge-on geometry. The comparison between the high- and zero-spin fits is shown for the PD and PA in panels (a) and (b) of Fig.~\ref{fig:kynbbrr_fit2}. The low-spin fit clearly fails to reproduce the increase of the PD in the highest energy bin. Since the fit was performed in Stokes space with the data binned into 30 bins, this poorer fit primarily reflects a worse reproduction of the $q$ and $u$ parameters at high energies.

\subsection{Dependence of reflected returning radiation on spin and albedo}

To better understand the strong spin dependence of the fit quality and polarization properties discussed above, we briefly summarize the expected physical behaviour of reflected returning radiation as a function of black-hole spin and albedo.

The polarization directions of the direct thermal radiation and the reflected self-irradiation differ significantly; when combined, one component partially depolarizes the other.

For black holes with low spin, the inner edge of the accretion disk lies further from the event horizon. As a result, only a small fraction of thermal photons return to the surface of the inner disk, while most of the returning photons fall below this edge. The lower temperature expected within $\sim\!10\,r_{\rm g}$ for a non-rotating black hole also implies weaker ionization, which further suppresses the reflected fraction. In this regime, the observed polarization is dominated by the direct thermal emission, and the weaker reflected returning radiation acts mainly to depolarize it. The resulting polarization direction is expected to be approximately perpendicular to the system axis.

In contrast, for rapidly spinning black holes ($a$ close to 1), the inner disk extends much closer to the event horizon. A large fraction of photons are bent back onto the disk and reflected, while the higher inner-disk temperatures increase ionization, further enhancing the reflection fraction. Consequently, the polarization becomes dominated by the reflected returning radiation, with the direct component acting as a depolarizing contribution. The resulting polarization direction is expected to be approximately parallel to the system axis.

The observed net polarization depends on the interplay between these two components, governed primarily by the spin and albedo, as well as by global system parameters such as the black-hole mass and accretion rate \citep[see, e.g.,][]{Schnittman2009}.


\bibliographystyle{aasjournalv7}

\bibliography{ref}

\end{document}